\documentclass{article}
\usepackage[utf8]{inputenc}
\usepackage{amssymb}
\usepackage{amsmath}
\usepackage{subfig}
\usepackage{graphicx}
\usepackage{todonotes}
\usepackage{authblk}
\usepackage{url}

\title{Statistical quality assessment of Ising-based annealer outputs}

\author[1]{Krzysztof Domino \thanks{kdomino@iitis.pl}}
\author[2]{M\'aty\'as Koniorczyk \thanks{koniorczyk.matyas@wigner.hu}}
\author[1,3]{Zbigniew~Puchała}

\affil[1]{Institute of Theoretical and Applied Informatics\\ 
	Polish Academy of Sciences\\ 
	Ba{\l}tycka~5, 44-100 Gliwice, Poland}
\affil[2]{Wigner Research Centre, H-1121 Budapest, Konkoly-Thege Mikl\'os \'ut 29-33, Hungary}
\affil[3]{Faculty of Physics, Astronomy and Applied Computer Science, Jagiellonian University, 30-348 Kraków, Poland}

\begin{document}

\maketitle

\section*{Abstract}

The ability to evaluate the outcomes of quantum annealers is essential for such devices to be used in complex computational tasks.
We introduce a statistical test of the quality of Ising-based annealers' output based on the data only, assessing the ground state's probability of being sampled. A higher probability value implies that at least the lower part of the spectrum is a part of the sample.
Assuming a plausible model of the univariate energy distribution of the sample, we express the ground-state energy and temperature as a function of cumulants up to the third order. Using the annealer samples, we evaluate this multiple times using Bootstrap resampling, resulting in an estimated histogram of ground-state energies and deduce the desired parameter on this basis. 
The approach provides an easily implementable method for the primary validation of Ising-based annealers' output. We demonstrate its behavior through experiments made with actual samples originating from quantum annealer devices.

\section*{Keywords}
Ising solver, quantum annealing, energy spectrum, quality assessment, cumulants

\section{Introduction}

Optimization problems have increasing importance in many fields~\cite{rosenhead2009reflections}, which is driven by several factors, including the demand for competitiveness, better use of resources, and the increasing complexity and interconnectivity in the contemporary world.
However, many problems of practical relevance are computationally hard~\cite{Aaronson13}~\cite{garey1979computers}. 
Quantum computational devices offer a promising perspective in handling such difficulties~\cite{Feynman60}. 
These include quantum annealers, such as the D-Wave machines~\cite{Lanting14}.
In principle, such machines could solve a variety of (hard) optimization 
problems ``naturally'' by finding low energy
eigenstates encoding the solution~\cite{harris_phase_2018,king_observation_2018}.
Therefore the development of quantum technology has the potential to efficiently solve 
complicated discrete optimization problems by encoding them into the energy of a physical system.
Consequently, offering the optimal solution as one of the ground states (since there exist Ising formulations of NP-hard problems~\cite{npising}). 
Indeed, during adiabatic evolution~\cite{Farhi00},  such a system reaches a ground state ``naturally''. Therefore in principle, an optimal solution of the encoded problem can be read out from the state deterministically. 

Quantum annealers such as the D-Wave machine are approximate physical realizations of an adiabatic quantum computer, i.e., they are based on a real physical system. They realize a fixed topology of couplings, e.g., the Chimera or Pegasus graphs. Thus the optimization problem must be embedded into this topology either as an induced subgraph or in a redundant manner using multiple physical quantum bits to represent a logical one. This procedure is called minor embedding and even though it is often doable relatively simply, finding an optimal embedding is a hard computational problem itself.

No real quantum system can be entirely separated from its environment: the phenomena such as the heat exchange~\cite{breuer_book} or decoherence~\cite{woj} cannot be wholly neglected in the case of a physical quantum annealer. This leads to a noisy version of the adiabatic evolution~\cite{venuti2016adiabaticity}. Moreover, the measurements performed at the last stage of computations are not perfect either. 
As the dimensionality of the underlying Hilbert space grows exponentially with the number of qubits~\cite{exp08}, the aforementioned issues affect the results to an even greater extent.
To tackle these problems in a quantum annealing device, the adiabatic evolution is run repeatedly, each run followed by a readout. The results form a statistical sample of configurations and objective values, possibly containing optimal solutions, i.e., the minimum energy states. 
The quantum annealer implements probabilistic heuristics, which are still potentially valuable for addressing specific hard computational problems.

The setting of the annealing procedure is challenging, even in the case of an ideal system. The parameters of the process depend on the minimum gap between the instantaneous ground state and the rest of the spectrum during the evolution, which in general, cannot be determined~\cite{Childs2001}. 
Therefore, in the observed result, which is the output from a physical quantum annealer, the elements of the output sample can have significantly higher energy than the ground states. Thus, whether any ground state has been sampled is an essential question in practical applications. 
In many cases, only the ground state is useful in physical applications, while in optimization problems, low-energy excited states are often also valuable. Therefore, obtaining states close to the ground state in the sample also bears significant relevance.

The estimation of the success probability of having a ground state (or states close to it in energy) in the sample, has also been addressed in benchmarking literature, see e.g.~\cite{willsch2020benchmarking,  willsch2021benchmarking, koshikawa2021benchmark}. In the state of the art benchmark scenarios, the addressed optimization problems are specifically designed, i.e. the ground state is usually known in advance, and the size of a problem is typically small. The success probability is then empirically investigated based on solving various benchmark problems and comparing the result with known solutions. Our idea is different: we propose the complementary method of estimating the \emph{ground-state energy} and testing the quality of the solution against containing any ground state solely using statistical analysis of the output of the Ising machine. We build on certain generic assumptions coming from the statistical description of the system, and therefore our method is more suitable for larger problems. This is in line with our intention to use it in application-driven scenarios. We will, however, use the minimum energy known from other calculations to validate our method.

Our goal is to use the entire sample resulting from a quantum annealer (or a similar device) to estimate the likelihood of having a ground state or at least states from the low-energy part of the spectrum in the sample. 
We remark here that as quadratic binary optimization problems are NP-hard, 
polynomial-time bounds on the optimal solution constitute important results in the classical literature on the problem, see e.g., the works of Nesterov~\cite{nesterov1998semidefinite} or Ye~\cite{Ye97approximatingquadratic}. Those results also provide hints to finding the distance between a particular solution and the optimal one. In a similar spirit, we can consider probabilistic solvers.

 In~\cite{albash2017temperature} it has been proposed that in future quantum annealer designs, to improve the convergence of solutions into the ground state with the increase of the number of spins, the temperature needs to be scaled down. For high enough temperatures, \cite{albash2017temperature}~assumes power-law scaling of the heat capacity with temperature. Such scaling is typically a near-phase-transition behavior, similar to that described by critical exponents. In~\cite{albash2017temperature} the relation between the first three cumulants of the Ising energy output and the heat capacity has been calculated using the Boltzmann distribution and the fluctuation-dissipation theorem. We will use this relation to estimate the ground state energy from the spectrum.

Our research is tied to Extreme Value Theory~\cite{coles2001introduction} where limits for low values are estimated from a particular probabilistic model of the sample. However, instead of estimating the extreme-value distribution, we assume its form from the underlying Ising model and use it to estimate the minimal values.

The paper is organized as follows. In Section~\ref{sec::theoretical} a theoretical introduction of our model is presented. In Section~\ref{sec::experiments} the results of our experiments using data both from simulators and real D-Wave machines are discussed. In Section~\ref{sec::conclusions} the results are summarised and conclusions are drawn. Appendix~\ref{sec:MH} contains more details on performing the Metropolis-Hastings simulations. 

\section{Theoretical model}\label{sec::theoretical}

Quantum annealers are based on the Ising model defined by the following Hamiltonian:
\begin{equation}\label{eq::Hamiltonian}
    H = \sum_{(i.j) \in \mathcal{E}} J_{i,j} s_i s_j + \sum_{i \in  \mathcal{V}} h_i s_i,
\end{equation}
where $\mathcal{V}$ is a set of spins (vertices), and $\mathcal{E}$ describes the topology of the processor. Furthermore, $s_i \in \{-1, 1\}$ is the spin value at the $i$'th vertex of $\mathcal{E}$, $J_{i,j}$ is the coupling between spin at vertex $i$ and spin at vertex $j$, and $h_i$ is the local field acting on spin at vertex $i$. 

In certain cases it is more convenient to deal with quadratic unconstrained binary optimization problems (QUBOs):
\begin{equation}
\min_{\mathbf{x}} \mathbf{x}^\top Q  \mathbf{x} \, ,
\label{eq::QUBO}
\end{equation}
where $\mathbf{x}$ is a binary vector of decision variables, and $Q$ is an arbitrary matrix
which can be, without loss of generality, symmetric or upper triangular.
There is a one-to-one relation~\cite{Glover2019} between Ising problems of finding the ground state of the Hamiltonian in Eq.~\eqref{eq::Hamiltonian} and the QUBO problems in Eq.~\eqref{eq::QUBO}, as their objective function's values depend linearly on each other. The physical quantum annealers are expected to reach a ground state of the classical Ising model, so even for a QUBO model, the energy samples will reflect Ising spectra. Therefore, our results will also apply to the output of the  QUBO formulations.

\subsection{Background}\label{sec::background}

Let us now briefly recapitulate the considerations in~\cite{albash2017temperature}, where the authors, using the techniques of statistical physics, analyze the energy spectrum of the Ising model under the assumption of the Boltzmann distribution. Concerning the effect of finite temperature, the analysis concludes that the probability of sampling the ground state goes to zero exponentially with the number of spins $N$. 
Thus, scaling down the temperature can help in regaining the success probability.

In the considerations of~\cite{albash2017temperature} there are two types of scaling of the specific heat with the temperature assumed: power-law for low values of $\beta = \frac{1}{T}$ and exponential for high values (albeit it is claimed as a general assumption that the system is not tuned to a phase transition point). We focus on the lower range, i.e. relatively high temperatures, as we expect our D-Wave samples to fall into this region. Hence, we will adopt the assumption that specific heat behaves as
\begin{equation}\label{eq::cscalling}
c(\beta) = - A \beta^{-\alpha -2},
\end{equation}
where $A$ is the coefficient of the particular instance, and $\alpha$ is a parameter of the model. The spread of energies is measured as a standard deviation of the sample of energies 
\begin{equation}
\sigma(H) = \sqrt{-N c(\beta)}.
\label{eq::sigma_H}
\end{equation}
Following~\cite{albash2017temperature} the mean of the energy distribution of the Ising system can be expressed as:
\begin{equation}\label{eq::Hstar_background}
    \langle H \rangle = E_0 -N \int_{\beta}^{\infty} c(\beta') d \beta' ,
\end{equation}
where $E_0$ is the ground state energy.
Following~\cite{albash2017temperature} asymmetry can be written as:
\begin{equation}\label{eq::skew_back}
   \eta(H) = \frac{1}{\sqrt{N}} \frac{1}{-c(\beta)^{3/2}} \frac{d c(\beta)}{d \beta}.
\end{equation}

Our model is based on assumptions expressed in  Eqs.~\eqref{eq::cscalling},~\eqref{eq::sigma_H},~\eqref{eq::Hstar_background}, and~\eqref{eq::skew_back}. Eq.~\eqref{eq::cscalling} describes the scaling, which is independent from the underlying distribution.
Remaining assumptions stem from our assumption of the Boltzmann distribution. 
In the case when a different distribution is assumed (as quantum annealer may also operate in the non-equilibrium region or even in the coherent one~\cite{amin2015searching}), these assumptions should be tested separately. In this way a different model version  can be obtained, which may outperform the original one if the assumed distribution is closer to the real one. The exact distribution is, however, unknown. Therefore, we assume the Boltzmann distribution and as shown later, with this assumption, the performance of our method is acceptable.

\subsection{Results}

From the assumptions presented in Section~\ref{sec::background}, we can derive relations, which will be used to asses validity of the Ising-based output. Using Eq.~\eqref{eq::sigma_H} (see also \cite{albash2017temperature}), we are given the formula for the variance:
\begin{equation}\label{eq::sigmaB}
\sigma^2(H) = N A \beta^{-\alpha -2}.
\end{equation}
Next, with the use of Eq.~\eqref{eq::sigmaB} and solving Eq.~\eqref{eq::Hstar_background}, we obtain:
\begin{equation}\label{eq::Hstar}
    \langle H \rangle = E_0 + N \frac{A}{\alpha + 1} \beta^{-\alpha - 1} = E_0 + \frac{\sigma^2(H) \beta}{\alpha + 1}.
\end{equation}
Analogously, by solving Eq.~\eqref{eq::skew_back}, we get
\begin{equation}\label{eq::skew}
   \eta(H) = \frac{1}{\sqrt{N A}} (\alpha + 2) \beta^{\alpha / 2} = \frac{\alpha + 2}{\sigma(H) \beta}.
\end{equation}
Finally combining Eq.~\eqref{eq::Hstar} and Eq.~\eqref{eq::skew} we obtain the formula for the ground state energy, which can be used as an estimator
\begin{equation}\label{eq::h_sigma}
    E_0 = \langle H \rangle - \frac{\alpha + 2}{\alpha + 1} \frac{\sigma (H)}{\eta(H)}.
\end{equation}
In the similar manner the parameter $\beta$ can be estimated 
\begin{equation}\label{eq::estimate_beta}
    \beta = \frac{E_0 - \langle H \rangle}{\sigma(H) \eta(H) (E_0 - \langle H \rangle) + \sigma^2(H)}.
\end{equation}
The energy of the ground state and the parameter $\beta$, under our assumptions, can be expressed as a function of cumulants $\sigma$, $\eta$ and $\langle H \rangle$, which will be later estimated from samples. Both in Eq.~\eqref{eq::h_sigma} and Eq.~\eqref{eq::estimate_beta} the asymmetry $\eta(H)$ can be computed as normalised $3$rd cumulant $c_3$ of the sample of energies, i.e.:
 \begin{equation}\label{eq::eta_from_cums}
 \eta(H) = \frac{c_3(H)}{\sigma(H)^3}.
 \end{equation}

\subsection{Estimation error analysis}

The output of the quantum annealer (or its simulator) is an $n$-sample of energies and configurations. Our method uses only the energies as input. The goal is to estimate the ground state energy using methods of moments, i.e., via cumulants computed from the data. Next, we will compare the estimate with the minimum value from the sample in order to assess the likelihood of the event that the sample contains the ground state energy. 

To construct the distribution of estimated ground-state energies $E_0$ in order to obtain a significance threshold, we use Bootstrap resampling~\cite{tibshirani1993introduction}. In details, let $H_1, H_2, \ldots , H_n$ be a sample, and $E_0(H_1, H_2, \ldots , H_n)$ be the estimate of the ground-state energy via Eq.~\eqref{eq::h_sigma}. 
Then from $H_1, H_2, \ldots  , H_n$ we sample $n$ items with replacements, i.e. repeating some of the elements optionally. Let us denote the resulting samples by
$H_1^{(j)}, H_2^{(j)}, \ldots  , H_n^{(j)}$. For each such sample we compute $E_0^{(j)}$. Repeating this procedure $S$ times we obtain the desired estimated distribution of $E_0$-s. 

To validate the Bootstrap approach, we compute the standard deviation of $E_0$  by $k$-statistics approximation and standard error calculus. In order to do so, it is convenient to combine Eq.~\eqref{eq::h_sigma} with Eq.~\eqref{eq::eta_from_cums}, then:
\begin{equation}
    E_0 = \langle H \rangle - \frac{\alpha + 2}{\alpha + 1} \frac{\sigma (H)^4}{c_3(H)}.
\end{equation}
Let $c_k$ be the non-normalised $k$-th order cumulant; we will omit the argument $H$ of all cumulants in what follows. 
The estimation error of $\langle H \rangle$ can be neglected in comparison to the estimation error of $\sigma^2$ and $c_3$, as the estimation error of moments (and cumulants) tends to increase with their degree.
To estimate the standard deviation of cumulants' estimation, we approximate the cumulants with $k$-statistics, which is valid for large $n$~\cite{weisstein2002k}. The standard deviation of the cumulants in the argument is:
\begin{equation}
 \delta c_3 \approx \sqrt{\frac{c_6 + 9 \sigma^2 c_4+ 9 c_3^2 + 6 \sigma^6}{n}},
\end{equation}
and
\begin{equation}
    \delta \sigma^2 \approx \sqrt{\frac{c_4 + 2 \sigma^4}{n}}.
\end{equation}
The contributions of the particular cumulants to the standard errors of $E_0$ are:
\begin{equation}\label{eq::e0_estim}
    \delta E_0(c_3) \approx \left|\frac{\partial E_0 }{\partial c_3}\right| \delta c_3, \ \  \delta E_0(\sigma^2) \approx \left|\frac{\partial E_0 }{\partial \sigma^2}\right| \delta \sigma^2.
\end{equation}
Finally, assuming that estimators are independent, the standard deviation of $E_0$ can be estimated as
\begin{equation}\label{eq::std_error}
    \delta E_0 \approx \sqrt{(\delta E_0(c_3))^2 + (\delta E_0(\sigma^2))^2}.
\end{equation}

\subsection{Physical assessment of validity}\label{sec::system_size}

The results in~\cite{albash2017temperature} serving as the basis for our considerations were obtained by considering and analyzing D-Wave solutions for the random instance with couplings $J = \pm 1$ as well as 3-regular 3-XORSAT instances and planted (droplet) instances on the Chimera graph.
In this context the assumption in Eq.~\eqref{eq::cscalling} results from the scaling $c(\beta) \sim T^{\alpha}$ which is appropriate for $\beta \neq \beta_c$ (such a scaling is a phase-transition-like behaviour in $\beta$), recall also that $\beta = 1/T$. Here $\beta_c$ is the quasi phase transition point. An actual phase transition would take place in a system of infinite volume. However, as discussed in~\cite{gligor2001econophysics}, for finite volume a phase-transition-like behaviour occurs instead. 
 The larger the system size, the sharper the dependence of $c(\beta)$ on $\beta$~\cite{malsagov2017analytical,binder2003overcoming}.
 We expect a real annealer to be in the $\beta < \beta_c$ regime as the thermal noise
 significantly impacts the machine's output. 

Consider a general probabilistic approach to the Ising model with couplings $J_{i,j}$-s and local fields $h_i$-s. However a phase-transition-like behavior can also be expected under such circumstances, in a less rigorous form. A simple model of variable coupling can be~\cite{kryzhanovsky2018investigation}
\begin{equation}
J_{i,j} = J_0 + \epsilon_{i,j},
\end{equation}
where $\epsilon$'s are drawn randomly according to a chosen probability distribution. In~\cite{kryzhanovsky2018investigation}, authors show, that the presence of the disturbance in such a form, indeed results in a flattening of the exponential scaling. One can claim that for small systems and high variation of couplings, $c(\beta)$ depends weekly on $\beta$.
Such a phenomenon can possibly affect, for example, the parameter $\alpha$, making it more instance-dependent in such a case.

Let us also remark that the heat capacity per node depends strongly on the number of connections of the node. In~\cite{albash2017temperature} authors assumed that the number of couplings scales linearly with $N$ and also, there were additional assumptions on the coupling strength. These do not necessarily hold for our problem instances. As both $\alpha$ and $\beta_c$ may vary with the degree of connectivity of the graph, the model may behave worse for a problem graph with a highly variable degree of connectivity. 

 The more detailed analysis of the parameter $\alpha$ will be a subject of further research: a more thorough investigation should include $\alpha$ parameter fitting for a particular type of instance and consider $\alpha$'s estimation error. Here, for demonstration, we will use the $\alpha = 0.19$ --- $0.38$ parameter values, already proposed in \cite{albash2017temperature}, as well as some lower value of $\alpha$ for wider sensitivity analysis.

\section{Experiments}\label{sec::experiments}

In what follows, we demonstrate the procedure introduced in the previous Section on particular examples. The inputs will be Ising annealer samples, and we shall test whether they contain the ground-state energy. 

The described method leads to the following algorithm:

\par\noindent{\bf Input:} 
\begin{itemize}
    \item Parameter $\alpha$.
    \item A sample of energies $\mathbb{H}= (H_1, H_2, \ldots, H_n)$ originating from $n$ independent runs of the Ising annealer.
\end{itemize}
\par\noindent{\bf Processing:} 
\begin{itemize}
    \item Step 1 (Initialization).
        \begin{itemize}
            \item  Define $H_{\min} = \min(H_1, H_2, \ldots, H_n)$;
        \end{itemize}
        
    \item Step 2 (Bootstrapping). \\ 
    Sample with replacements from $\mathbb{H}$ to obtain $S$ Bootstrap samples 
    $$ \mathbb{H}^{(j)} = (H^{(j)}_1, H^{(j)}_2, \ldots, H^{(j)}_n),$$ 
    where $j = 1, \dots, S$;
    
    \item Step 3 (Estimation of the conditional minimum). \\ 
    For each $\mathbb{H}^{(j)}$ estimate the minimal value $E_0^{(j)}$ from the first 3 momenta using Eq.~\eqref{eq::h_sigma} with the given value of the parameter $\alpha$;
    
    \item Step 4 (Estimation of the conditional distribution). \\ 
    Define a conditional probability measure $\mu^{(\mathbb{H})}$ as an empirical distribution of the data $\left\{E_0^{(j)} \right\}_{j=1}^S$;
    
    \item Step 5 (Conditional likelihood). \\ 
    Calculate $p$-value, i.e.
    $$
    p_{\mathrm{val}} = P_{\mu^{(\mathbb{H})}}(X > H_{\min}) =  1 - P_{\mu^{(\mathbb{H})}}(X \leq H_{\min}),
    $$
    where $X$ is a random variable with the distribution $\mu^{(\mathbb{H})}$;
    
\end{itemize}
\par\noindent{\bf Output:}
\begin{itemize}
    \item $p_{\mathrm{val}}$
\end{itemize}
The above procedure provides an estimate of the probability of having a ground state in the sample. A higher parameter value also implies probabilistically that at least a lower part of the energy spectrum has been sampled.
The estimation in Step 3 assumes the theoretical model from Section~\ref{sec::theoretical} and 
 depends on the original sample ($\mathbb{H}$) and on the result of the draw from Step 2.
This estimated probability $p_{\mathrm{val}}$ of having sampled a ground state will be referred to as the "$p$-value" in what follows. 
The algorithm was implemented in Julia programming language, and the source code is publicly available~\cite{statisticalcode}.

\subsection{Artificial data}
\label{sec:artificial}

Our first experiments were performed on a problem instance of $198$ logical bits from the field of railway operations research, described in detail as a case $1$ example in~\cite{domino2020quantum} (consult Section IV A therein for a problem description and Section III C for its QUBO formulation). 

\begin{figure}
    \centering
    \subfloat[]{
    \includegraphics[width=0.45\textwidth]{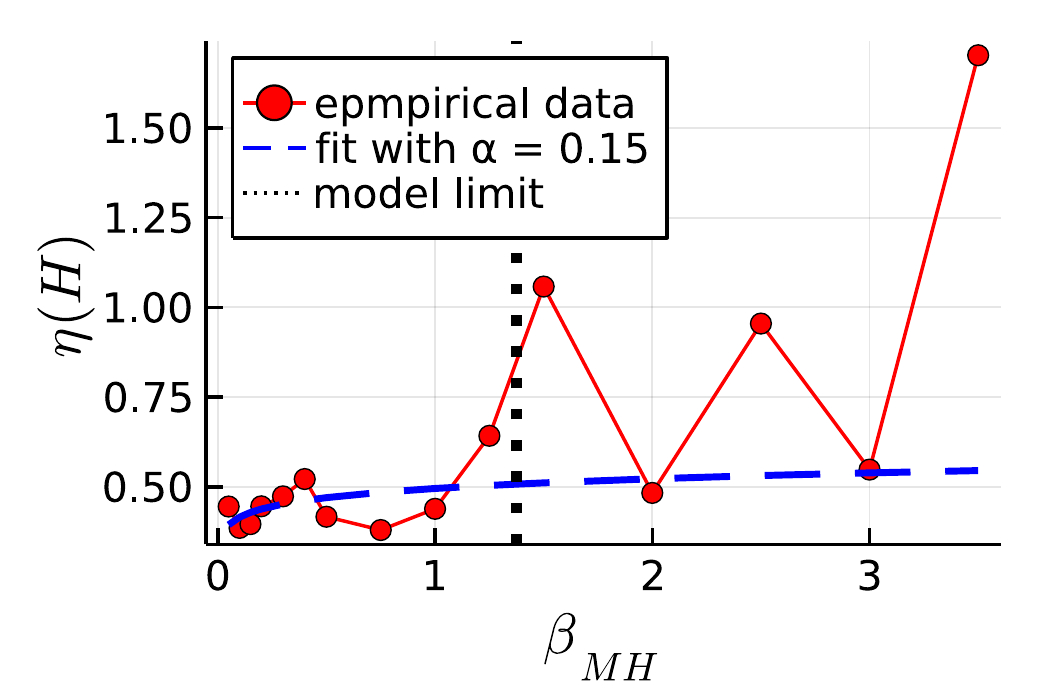}}
    \subfloat[]{
    \includegraphics[width=0.45\textwidth]{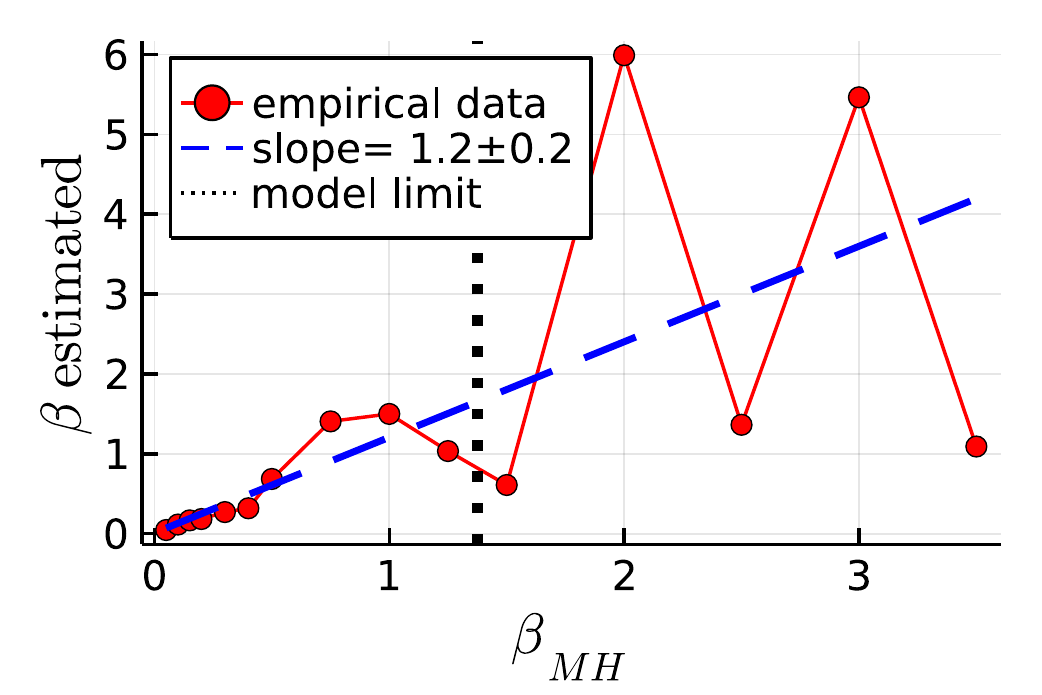}\label{fig::betafit}}
    \caption{Output of Metropolis-Hastings sampling with $n = 1000$ samples. The asymmetry $\eta(H)$ (left panel) and the estimated $\beta$ (right panel), both are functions of the parameter $\beta_{\text{MH}}$ of the Metropolis-Hastings algorithm. In left panel the following fitting model  has been used: $\eta \propto \beta_{\text{MH}}^{\alpha/2}$, see Eq.~\eqref{eq::skew}; fitting has been performed up to the threshold $\beta_{MH} = 1.375$, termed as the model limit. The limit has been determined by the visual analysis of $\eta(H)$ fitting it to the plot. (Recall that the limit can be case dependent). Observe that $\beta_{MH} \approx \beta$ within model validity, while accounting for the error margin. This validates our model.}
    \label{fig::betas}
\end{figure}

The samples have been generated using the Metropolis-Hastings algorithm. We refer to these data as artificial as they do not come from a physical solver.
The Metropolis-Hastings algorithm has $\beta_{\text{MH}}$ as a parameter playing a similar role as the $\beta$ in our model in the range of model validity. It is tied to the temperature of the simulated system and thus affects the quality of the solution. We expect $\beta_{\text{MH}} \approx \beta$; the latter estimated in the way described in Section~\ref{sec:betaestim}. The results on $\beta$ estimation and the scaling of $\eta(H)$ vs. $\beta_{MH}$, see Eq.~\eqref{eq::skew}, are presented in Fig.~\ref{fig::betas}. As it was expected in Section~\ref{sec::system_size}, we have a limit on $\beta_{\text{MH}}$, below which scaling $\eta \propto \beta_{\text{MH}}^{\alpha/2}$, and Eq.~\eqref{eq::cscalling} hold. To find the threshold, we can analyze the scaling of $\eta(H)$ and determine the threshold value of $\beta_{MH}$ above which it stops following the model in Eq.~\eqref{eq::skew}. It has been done manually in the case of Fig.~\ref{fig::betas} (left panel), but it could be done automatically.

\begin{figure}
    \centering
    \includegraphics[width=0.65\textwidth]{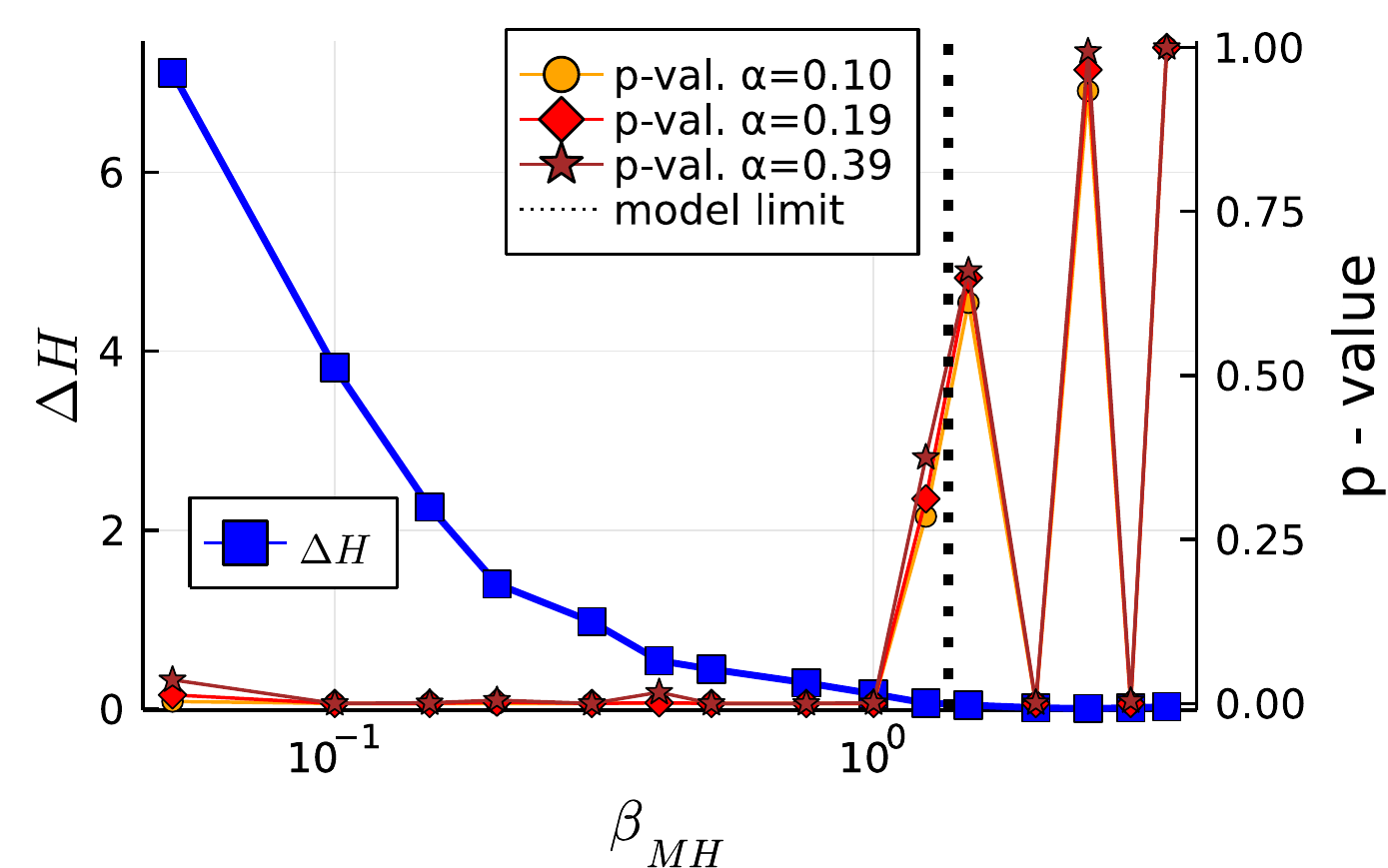}
    \caption{The minimal energy and the $p$-value for the problem instance addressed in Section~\ref{sec:artificial}. We have used $S = 1000$ for bootstrapping. In the model validity region $\beta_{MH} < 1.375$, we can observe $9$ solutions that are far from the ground state; those have an almost zero $p$-value and one solution that is near the ground state with the $p$-value of $\approx 0.3$. This is as expected. The choice of the $\alpha$ values is discussed at the end of Section~\ref{sec::system_size}. Observe that the particular choice of the value of parameter $\alpha$ does not significantly affect the results. Note also that there was no minor embedding in this case. We have used $n = 1000$ samples from the Metropolis-Hastings algorithm's output.}
    \label{fig::p_val_artificial}
\end{figure}

\begin{figure}
    \centering
    \includegraphics[width=0.45\textwidth]{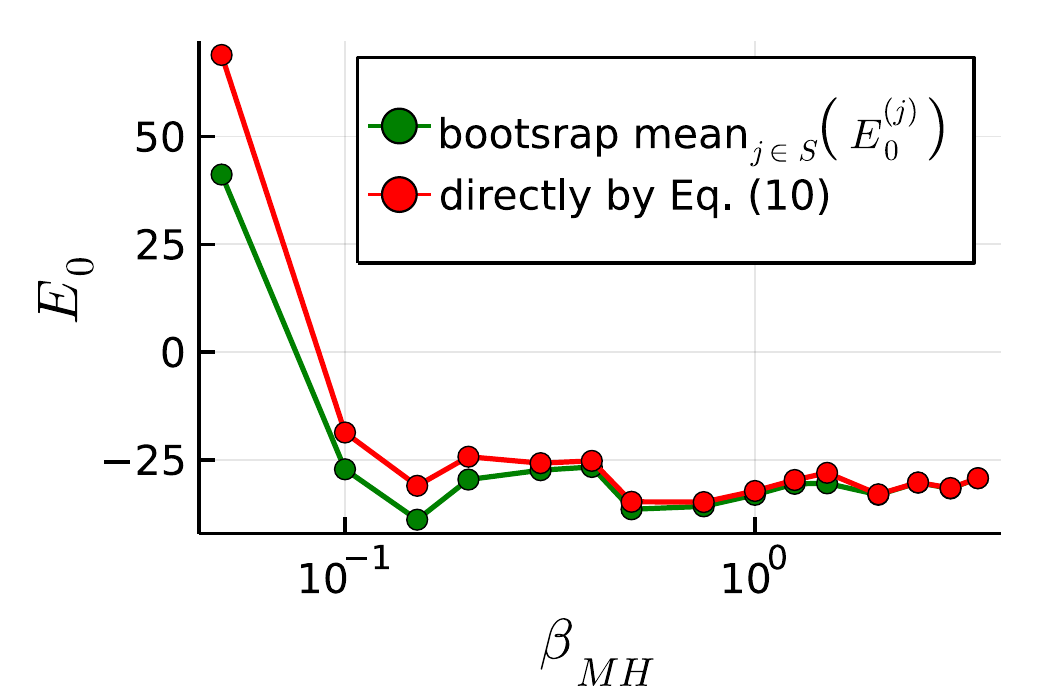}
    \includegraphics[width=0.45\textwidth]{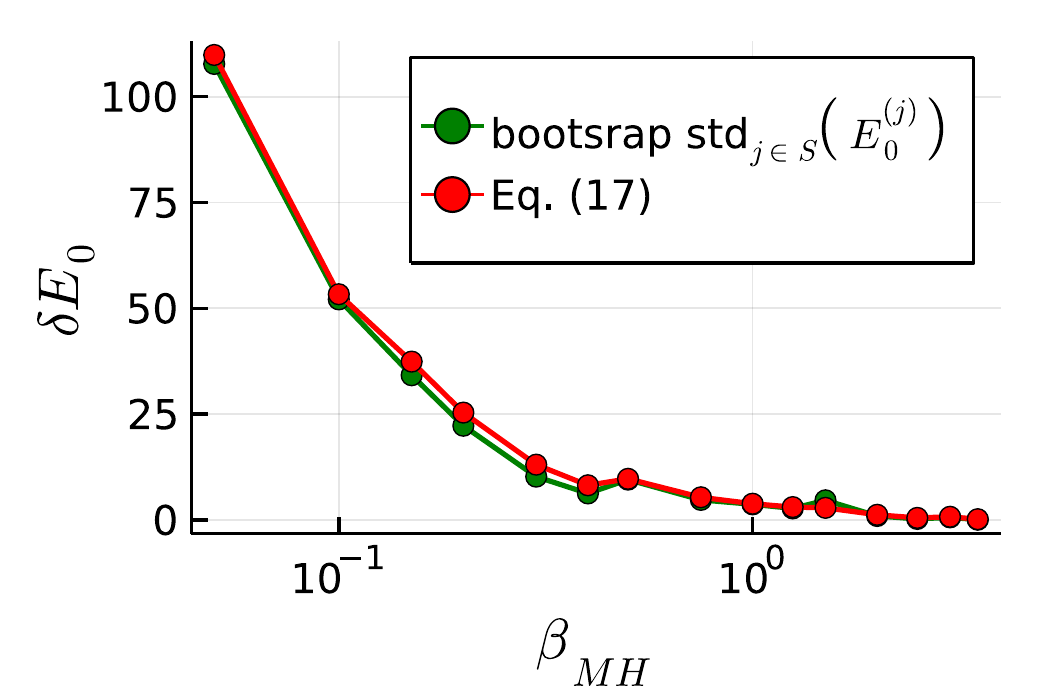}
    \caption{The comparison of $E_0$ obtained using Eq.~\eqref{eq::h_sigma} and $\sigma(E_0)$ from ~Eq.~\eqref{eq::std_error}, with mean and standard deviation of $\{E_0^{(j)}\}_{j = 1}^S$ for the Bootstrap sample, with $\alpha = 0.19$. In both cases the series are similar, validating the Bootstrap approach for the analysed data.
    }
    \label{fig::hists}
\end{figure}

In the case of this problem instance, the optimum is known. Hence, we can plot both the difference of the minimum energy state in the sample $H_{\min}$ from the ground state $E_0$, and also the $p$-value as a function of $\beta_{\text{MH}}$. 
To assess the quality of solutions, we use the relative difference between the best solution $H_{\min}$ and the ground state $E_0$:
\begin{equation}\label{eq::h0}
    \Delta H = \frac{H_{\min} - E_0}{| E_0 |},
\end{equation} 
which is equal to zero only if the ground state energy value is in the sample. (The division by $|E_0|$ can be omitted if it is more convenient to check the absolute differences.)
Fig.~\ref{fig::p_val_artificial} clearly demonstrates that for $\beta_{\text{MH}}$-s within the model validity region, the $p$-value can be used to distinguish between better and worse solutions. 
Finally, in Fig.~\ref{fig::hists} we have compared the means and the standard deviations of the Bootstrap histograms and found them to coincide with those obtained by the direct calculation (with no bootstrapping) via Eq.~\eqref{eq::h_sigma} and Eq.~\eqref{eq::std_error}.

\subsection{Analysis on D-Wave data}

We have tested our algorithm on energy spectra returned by a D-Wave quantum annealer. The addressed optimization problems were the aforementioned set of practical instances as well as the a set of droplet instances. As an additional set of examples, we present results on randomly generated exact set cover problems which we will be described later. 

The first set contains instances from the field of railway operations research~\cite{domino2020quantum}, see Section IV therein for a more detailed description. The second set contains droplet instances characterized by artificially "planted" ground states~\cite{rams2018heuristic}, designed to be difficult for an annealer. 
Droplets are specially designed to benchmark various annealers; they are not motivated by other optimization problems.
Note that the two types of instances differ in the sense of the variability of couplings and local fields (this issue is important as discussed in Section~\ref{sec::system_size}). In the case of practical instances, there are many couplings with the same value reflecting a particular set of constraints in the actual problem, which does not hold for the droplet instances. 

\begin{figure}
    \centering
        \subfloat[]{
    \includegraphics[width=0.47\textwidth]{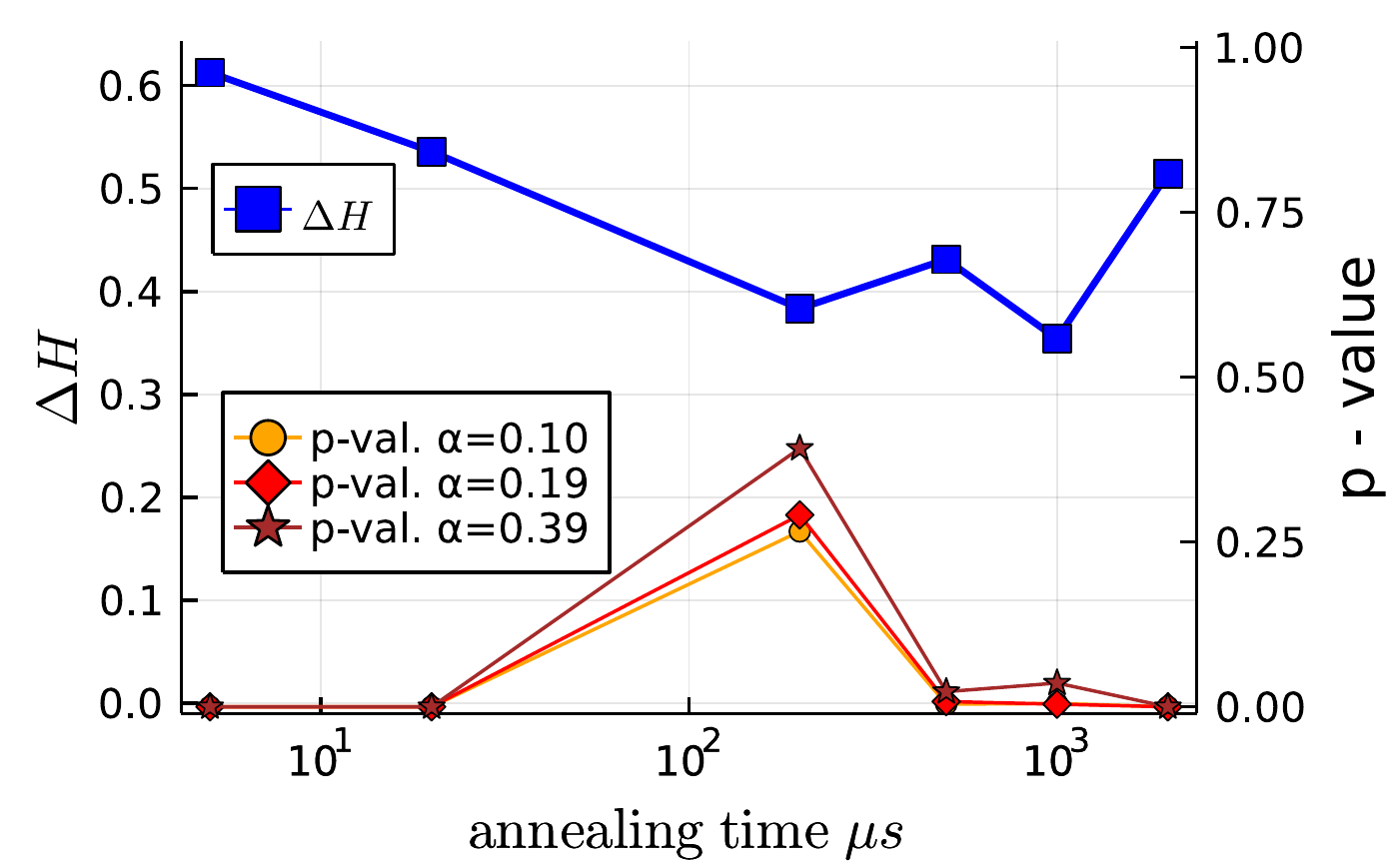}\label{fig::p_val_chimera_trains}
    }
    \subfloat[]{
    \includegraphics[width=0.47\textwidth]{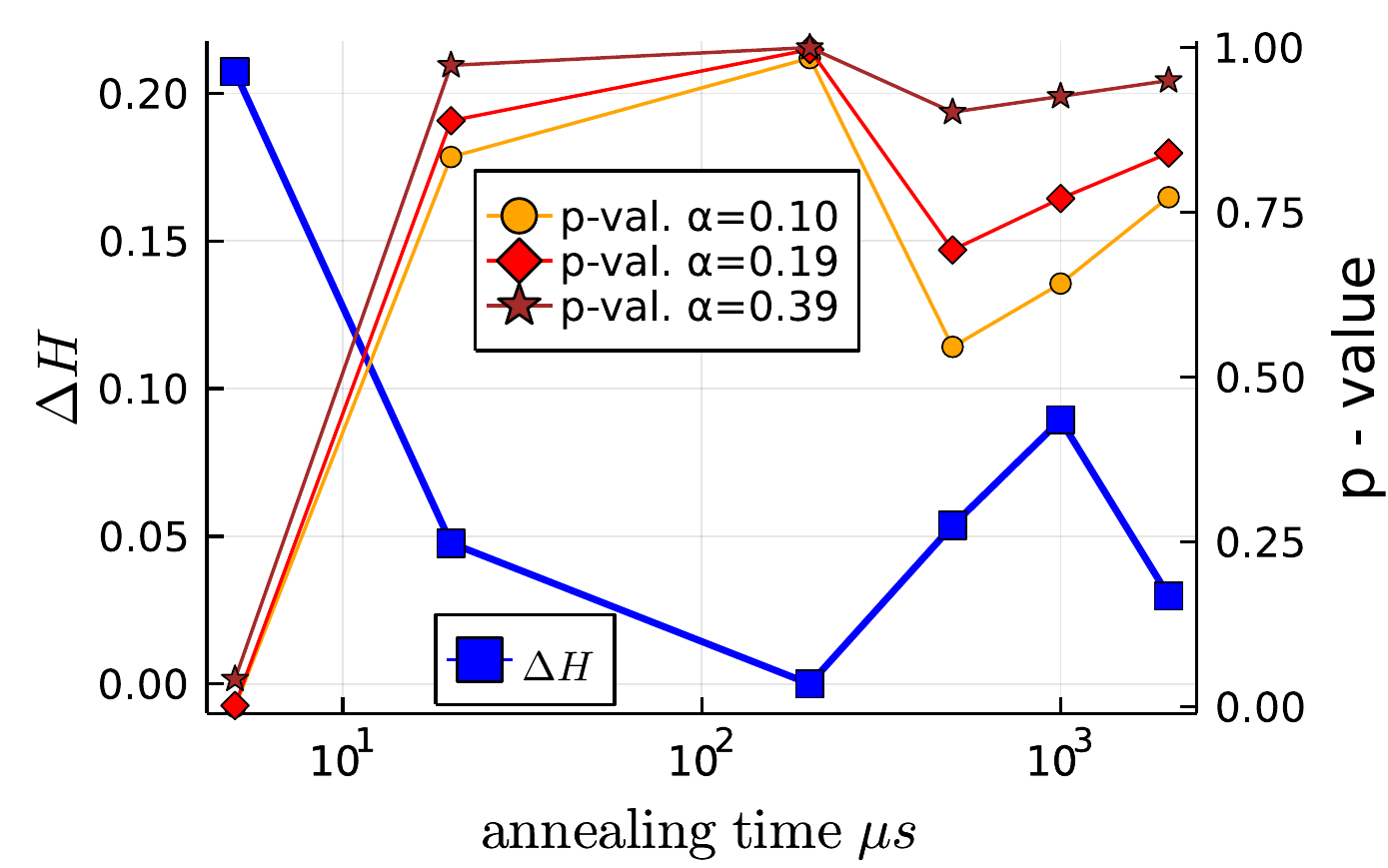}\label{fig::p_val_chimera_small_trains}
    }
    \caption{The $p$-value validation of the D-Wave (QPU: DW\_2000Q\_6) solutions of the Chimera practical instances from~\cite{domino2020quantum} (Section IV C therein). The dependence of the $p$ value and $\Delta H$, the relative difference between the best D-Wave solution and the actual minimum (c.f. Eq.~\eqref{eq::h0}) is plotted against the annealing time. The problem sizes are left panel: $198$ logical bits, right panel: $48$ logical bits. Chimera minor embedding was used, as described in~\cite{domino2020quantum}, with coupling strength $css = 2$; $n = 1000$ D-Wave samples were involved. In most cases, the higher $p$-value reflects the lower $\Delta H$. 
    }\label{fig::p_val_trains_chim}
\end{figure}

\begin{figure}
    \centering
    \includegraphics[width=0.65\textwidth]{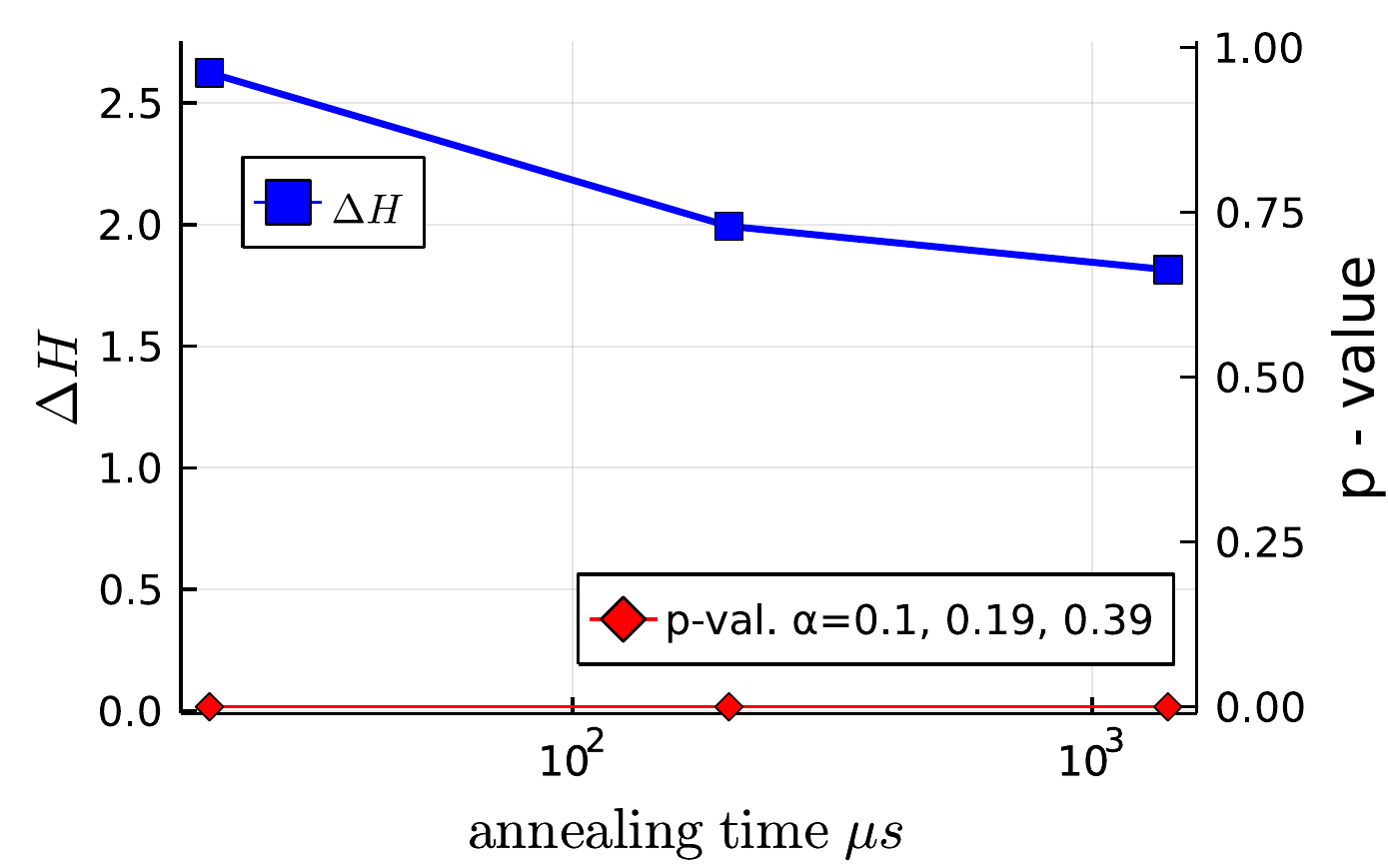}
    \caption{The $p$-value validation of D-Wave (QPU: Advantage\_system1.1) Pegasus instance from~\cite{domino2020quantum} with $594$ logical bits: The dependence of the $p$ value and the relative difference $\Delta H$ in Eq.~\eqref{eq::h0} is plotted against the annealing time. Pegasus embedding and $n = 25 000$ D-Wave samples were used. The results' high $\Delta H$ 
    is reflected in the zero $p$-values.}\label{fig::p_val_trains_pegasus}
\end{figure}

Data sets are returned by two D-Wave quantum annealers: the D-Wave $2000$Q (Chimera based); see Figs.~\ref{fig::p_val_chimera_trains},~\ref{fig::p_val_chimera_small_trains}, Tabs.~\ref{tab::p_val_chimera_droplet},~\ref{tab::e_val_chimera_droplet}, Tabs.~\ref{tab::p_val_chimera_droplet_small},~\ref{tab::e_val_chimera_droplet_small} and the D-Wave $5000$ ``Advantage System'' (Pegasus based); see Fig.~\ref{fig::p_val_trains_pegasus}. In all the experiments, we have used $S = 1000$ for bootstrapping. 
The quality of the samples depends on the annealing time, hence, it is important to decide whether it was chosen appropriately. 

In the case of the practical instances coming from~\cite{domino2020quantum} we, therefore,
plot our $p$-values along with the difference of the energy from the true ground state, see Eq.~\eqref{eq::h0}, as a function of the annealing time. This is done in order to demonstrate the potential usefulness of the $p$-value in the estimation of whether the right annealing time had been chosen. (We will return to the estimation of the $\beta$ parameter later.) The figures Fig.~\ref{fig::p_val_chimera_trains}, Fig.~\ref{fig::p_val_chimera_small_trains}, and Fig.~\ref{fig::p_val_trains_pegasus} all confirm that the $p$-value has the expected behavior: in most cases it reflects whether the best solution from the sample is close to the ground-state energy, as it can be expected from a probabilistic discriminator. The useful solutions: those near the ground state (i.e. whose energy is within $10 \%$ reach of the minimum, see Fig.~\ref{fig::p_val_chimera_small_trains}) give high $p$-values (i.e. above $0.5$), whereas solutions far from the ground state yield low $p$-values. 
For the set of our practical instances solved on D-Wave, the usefulness of the method has thus been demonstrated. 

The results for the droplet instances of $2048$ physical quantum bits are presented in Tabs.~\ref{tab::p_val_chimera_droplet},~\ref{tab::e_val_chimera_droplet}, the results for small droplet instances of $128$ physical quantum bits are presented in Tabs.~\ref{tab::p_val_chimera_droplet_small},~\ref{tab::e_val_chimera_droplet_small}.  
As one can observe, for some droplet instances and experimental settings, the model works well (black numbers), and for some of them (red numbers) it does not. In Tab.~\ref{tab::p_val_chimera_droplet} there are points where $p$-value based test concludes positively: the high value suggests that the ground state should have been sampled, even though it is not the case. Therefore we consider these as "false positives".  In Tab.~\ref{tab::p_val_chimera_droplet_small} we have  "false negatives", these are points where $p$-value based test falsely gives a negative results; it incorrectly indicates the absence of the ground state from the sample. It will be shown in~Section~\ref{sec:betaestim} that most of the false (red) results can be filtered out by analysing the estimated temperature of the annealer, as such an estimate will not be physical for most of problematic data. 

\begin{table}[ht]
\centering
\subfloat[$p$-values of particular instances with different annealing times]{
\begin{tabular}{lcccc}
\hline
  \multicolumn{4}{c}{\textbf{$p$-value}} \\ \hline
\textbf{}     & it. $1$  & it. $2$   & it. $3$  & it. $4$  \\ \hline
$5 \mu$s & {\color{red} $0.448$} & $0$  & $0$ & {\color{red} $0.128$}             \\ \hline
$20 \mu$s & $0$ & $0.001$  & $0.009$ & $0.001$              \\ \hline
$200 \mu$s & $0.036$ & $0$  & $0$ & $0$                \\ \hline
$2000 \mu$s & $0.002$ & $0$  & $0$ & $0.008$           \\ \hline
\end{tabular}\label{tab::p_val_chimera_droplet}
}
\quad
\subfloat[Absolute difference between the best D-Wave sample and the ground state energies; the real ground state energy is approx $-3400$]{
\begin{tabular}{lcccc}
\hline
  \multicolumn{4}{c}{$H_{\min} - E_0$} \\ \hline
\textbf{}      & \textbf{$1$}  & \textbf{$2$}    & \textbf{$3$}   & \textbf{$4$}  \\ \hline
$5 \mu$s & $98$ & $106$  & $120$ & $96$              \\ \hline
$20 \mu$s & $83$ & $96$  & $102$ & $90$              \\ \hline
$200 \mu$s & $64$ & $81$  & $76$ & $61$  \\ \hline
$2000 \mu$s & $45$ & $67$ & $64$ & $45$  \\ \hline
\end{tabular}\label{tab::e_val_chimera_droplet}
}
\caption{The $p$-value evaluation of solutions for a droplet instance of  $2048$ physical quantum bits ($\alpha = 0.19$). We used $n = 2 500$ D-Wave samples. An item index is an ordinal number. Such instances are designed to be difficult for annealers, which is reflected in the low $p$-values. False positives cases, i.e., where the $p$-value is high for a non-optimal solution, are marked red. Note that the estimation of $\beta$ resulted in a nonphysical value in these cases, which indicates an issue with the model's validity.}
\label{tab::falsepositives}
\end{table}

\begin{table}[ht]
\centering
\subfloat[$p$-values of particular instances for different annealing times]{
\begin{tabular}{lcccc}
\hline
  \multicolumn{4}{c}{\textbf{$p$-value}} \\ \hline
\textbf{}     & it. $1$  & it. $2$   & it. $3$  & it. $4$  \\ \hline
$5 \mu$s & $0.160$ & $0$  & $0.061$ & $0$              \\ \hline
$20 \mu$s & $0.265$ & $0$  & {\color{red} $0$} & {\color{red} $0.004$}                \\ \hline
$200 \mu$s & $0.664$ & {\color{red} $0.002$}  & {\color{red} $0$} & {\color{red} $0.001$}           \\ \hline
\end{tabular}\label{tab::p_val_chimera_droplet_small}
}
\quad
\subfloat[Absolute difference between the best D-Wave sample and the ground state energy of approx. $-210$]{
\begin{tabular}{lcccc}
\hline
  \multicolumn{4}{c}{$H_{\min} - E_0$} \\ \hline
\textbf{}      & \textbf{$1$}  & \textbf{$2$}    & \textbf{$3$}   & \textbf{$4$}  \\ \hline
$5 \mu s$ & $0$ & $0.24$  & $0$ & $1.15$              \\ \hline
$20 \mu s$ & $0$ & $0.21$  & $0$ & $0$  \\ \hline
$200  \mu s$ & $0$ & $0$ & $0$ & $0$  \\ \hline
\end{tabular}\label{tab::e_val_chimera_droplet_small}
}
\caption{The $p$-value valuation of solutions for a droplet instance of $128$ physical quantum bits ($\alpha = 0.19$). We used $n = 1000$ D-Wave samples. An item index is an ordinal number. 
False negatives, real optima with a low $p$-value, are typeset in red. Similarly to the case of the false positives presented in Table~\ref{tab::falsepositives}, the $\beta$ estimation fails in some of these cases, which indicates the model validity issue.}
\end{table}

From the above results, one can conclude that, in general, if the output is far from the ground state (e.g., see the worst results on each figure or table), we have almost zero $p$-value. Hence, a high $p$-value can be a valid indicator indicating that the solution is in the low energy part of the spectrum. (Recall, however that in the case of droplet instances, we are interested only in the ground state, whereas low excited states can also be of interest in some other problems.)
We can consider such a procedure as a \emph{primary valuation} of the solution.

As we have seen in the examples, our discriminator will yield both false positives and false negatives. Their presence can be due to various reasons, including the deviation from the Boltzmann distribution, failure of the scaling in Eq.~\eqref{eq::cscalling} for some instances, 
the error of the estimation of the third-order momenta, the use of inappropriate $\alpha$ parameter. Also, as most problems need to be embedded in the annealer's graph, our method could be best applied to the embedded problem's raw data. The minimal postprocessing to get the solutions to the original problem from the embedded data (including also the fixing of chain breaks with majority voting) can also affect the accuracy of our method. As in practice, one would prefer working with the original (and not the embedded) problems, we have tested our method in this way, and it appears to work also in this setting.

As an additional test, we have addressed set partitioning problems. 
Set partitioning and cover problems constitute a relevant class of hard optimization problems, with significant application in staff scheduling problems~\cite{Ernst2004}. Their large instances require specialised algorithms (see~\cite{Tahir2019} for a recent study). Even though the small ones can be solved using a linear constrained 0-1 program formulation easily, they have been used to benchmark quantum annealers recently~\cite{willsch2021benchmarking}, as they are well-known and controllable problems.
We have generated random exact cover problems of small sizes and have solved them in the linear 0-1 problem formulation. Then we have converted them into a QUBO form using penalties as described, e.g. in ~\cite{Glover2019}, choosing a large enough penalty coefficient. This procedure resembles a typical practical application. We have solved the so arising QUBO using its standard linearization~\cite{fortetstdlin}, resulting in a relatively large mixed-integer program that we have solved with GLPK~\cite{GLPK202203213604}. This way, we were aware of the minimum of the QUBO. We have verified that the solution of the QUBO coincides with the classical solution, so our choice of the penalty was correct. From this point on, we are only interested in analysing QUBOs: if the minimum will get actually sampled when solved on a physical annealer.

\begin{figure}
    \centering
    \includegraphics[width=0.65\textwidth]{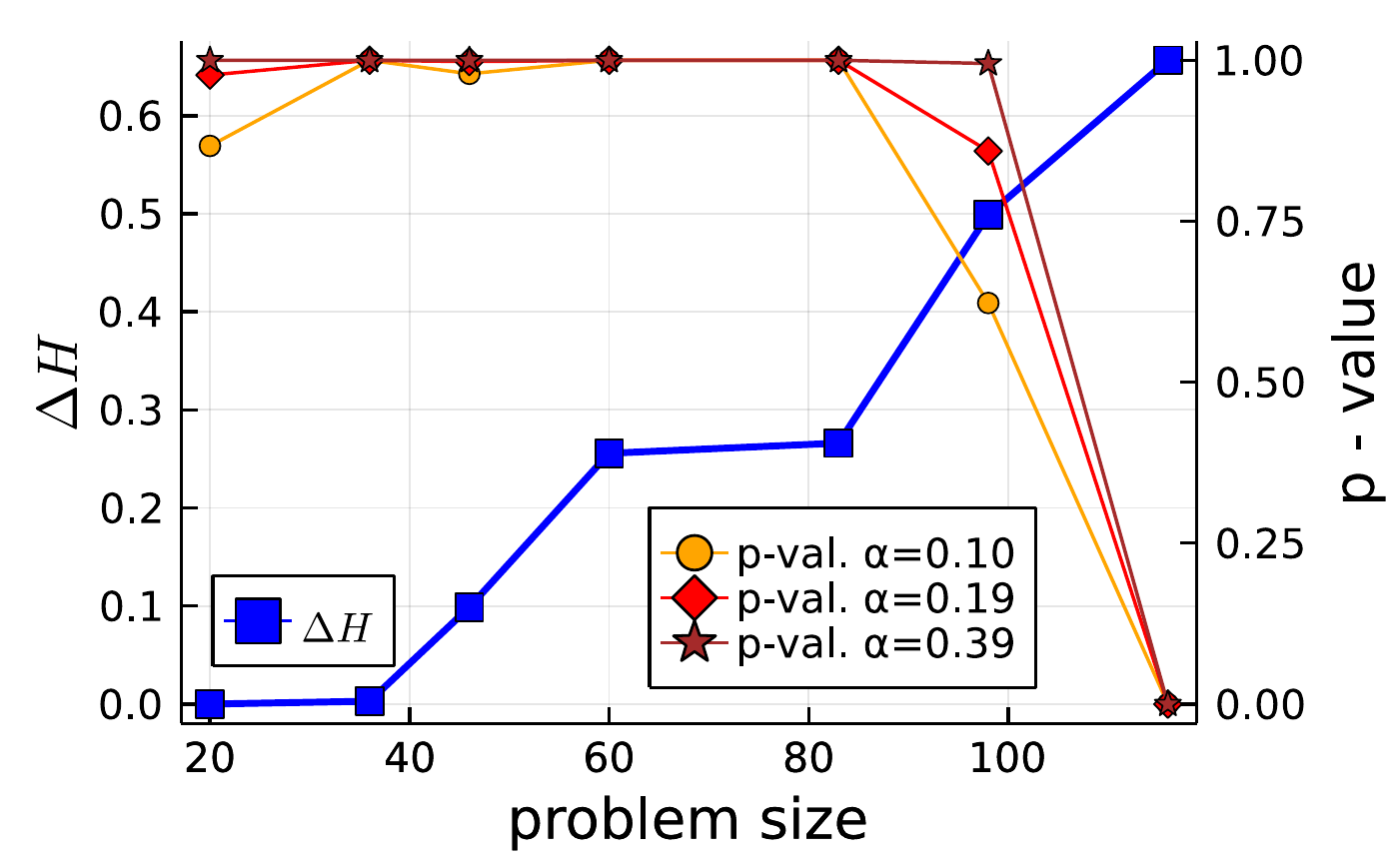}
    \caption{Exact cover problem instances. The dependence of the $p$-value and the relative difference $\Delta H$ in Eq.~\eqref{eq::h0} is plotted against the problem size, i.e. the number of binary decision variables. We have obtained $n = 3~996$ D-Wave samples at $250 \mu s$ of annealing time, with auto-scaling, embedding, and the default (minimal) postprocessing. The $p$-value analysis provides valuable information in most cases, despite the possible departures from its base assumption.}
    \label{fig::setpart}
    \end{figure}

We have solved the problem instances on the D-Wave Advantage4.1 quantum computer. To keep ourselves in a situation similar to practice, we have used the default settings of autoscaling, embedding, and minimal postprocessing. We have applied our method to the so-obtained samples.
The results are presented in Figure~\ref{fig::setpart}. The true minimum was sampled only in the case of the smallest instance, and we are getting farther in the case of the bigger problem instances. This illustrates further that our method can be useful, at least to some extent, even though, in the case of these calculations, our original assumptions on the distribution do not probably hold exactly because of the use of QUBO formulation, the embedding, and the auto-scaling and minimal postprocessing.

\subsection{Beta estimation}
\label{sec:betaestim}

If the ground state energy is known with certainty (e.g., from some other consideration),
we can estimate the parameter $\beta$  by means of Eq.~\eqref{eq::estimate_beta}. This calculation of $\beta$ can be used to validate the effectiveness of our method on a particular set of instances. The $\beta$ calculation requires the ground state energy for a few instances within the set of problems calculated under the same circumstances. To demonstrate such an approach conceptually, we compare $\beta$ computed for droplet instances and practical instances. Alternatively, Eq.~\eqref{eq::estimate_beta} can be reformulated to express $\beta$ in terms of $\alpha$ rather than $E_0$, as a result, more instance-wise validation will be possible.

The $\beta$ parameter reflects the effective temperature of the Ising system realised by the annealer. Therefore this parameter carries information about the extent of noisiness to expect: the higher the $\beta$, the closer the annealer is to an ideal adiabatic quantum computer. (Meanwhile, as seen in Fig.~\ref{fig::betas}, the estimation of $\beta$ facilitates the comparison with the Metropolis-Hastings approach.)

\begin{figure}
    \centering
    \subfloat[D-Wave Chimera $48$ logical bits]{
      \includegraphics[width=0.45\textwidth]{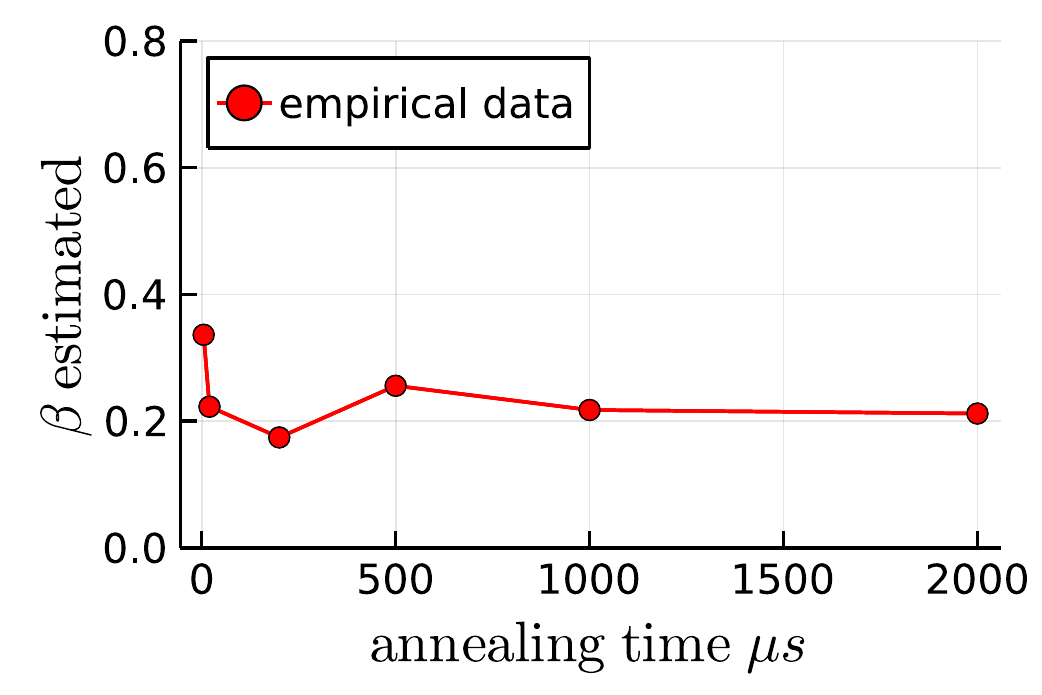}}
    \subfloat[D-Wave Chimera $198$ logical bits]{
      \includegraphics[width=0.45\textwidth]{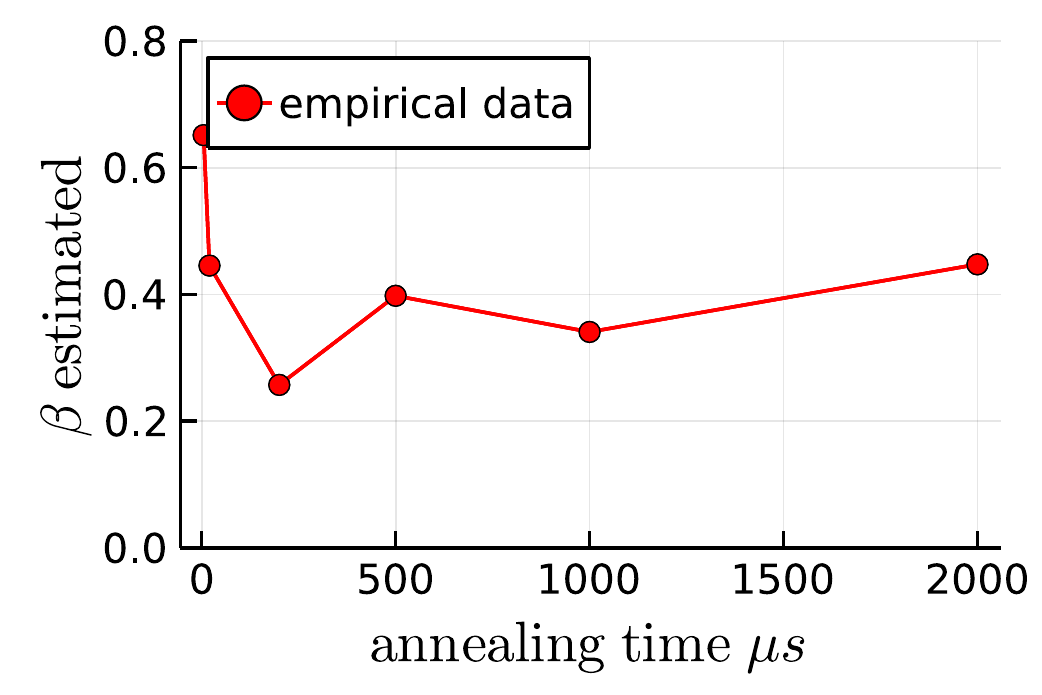}} \\
      \subfloat[D-Wave Pegasus solver $594$ logical bits]{
  \includegraphics[width=0.45\textwidth]{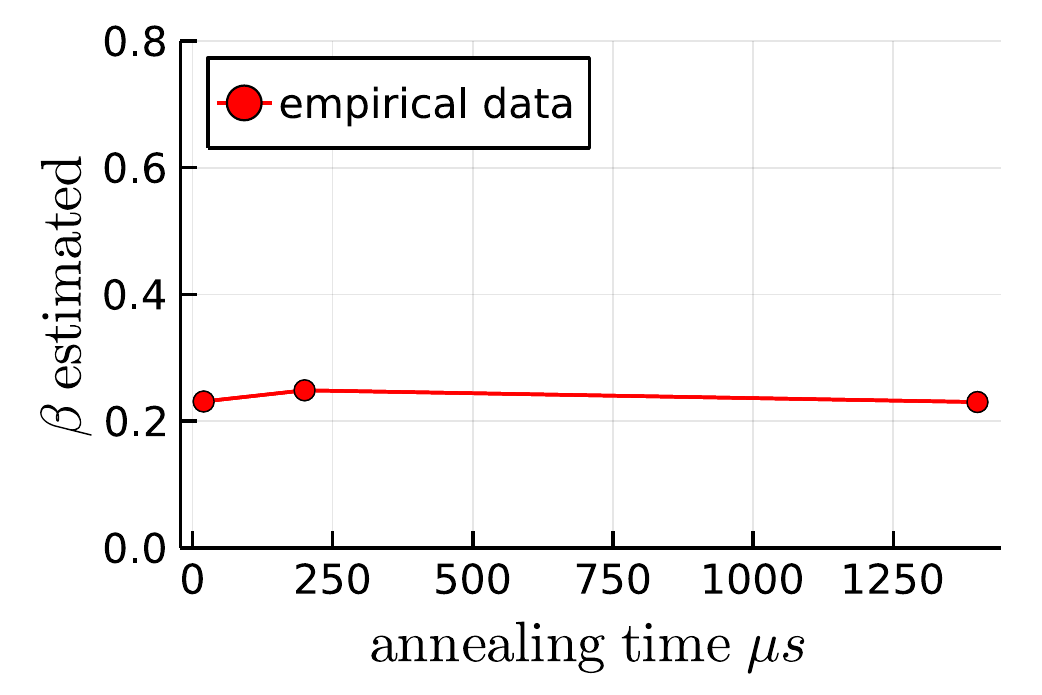}}
    \caption{$\beta$ estimation from practical instances on the D-Wave Chimera and Pegasus (specifics as in Fig.~\ref{fig::p_val_trains_chim} and Fig.~\ref{fig::p_val_trains_pegasus}). Observe that the $\beta$ values appear to be physical (i.e. positive) for all experimental points. However, in these cases, the temperature was computed from the QUBO energy spectrum (not Ising one), and the auto-scale was on. Hence it is re-scaled in comparison with the actual annealer temperature.}
    \label{fig::beta_chimera_trains}
\end{figure}

The results of the $\beta$ computation are presented in Tabs.~\ref{tab::beta_droplet},~\ref{tab::beta_droplet_small} (droplet) and Fig.~\ref{fig::beta_chimera_trains} (practical instances).
There is a large spread in $\beta$ estimation for the droplet instances. Furthermore, some values of $\beta$ are non-physical (i.e. negative). The observed behavior can be caused by the fact that droplets are complicated instances with highly variable couplings. Hence our model may not work correctly for some (especially smaller) droplet instances. For such smaller instances, the physical validity of the model may be weaker, see Section~\ref{sec::system_size}). Here, however, a very important conclusion appears. If we have a series of instances (e.g. droplets), and  we know the ground state exactly for a few, we can check if the $\beta$ values are physical for these few. Based on this, we can conclude whether the method works well or not for the whole series.

In the case of practical instances, where the instance itself is less complicated, estimated $\beta$ values appear to be physical for all data, see Fig.~\ref{fig::beta_chimera_trains}. Furthermore, for the particular Chimera chip, $\beta$ oscillates in the range $0.2$ - $0.65$. In contrast, for the particular Pegasus chip, it is closer to $0.2$, which may suggest a slightly higher temperature on the Pegasus chip.

\begin{table}[ht]
\centering
\begin{tabular}{lcccc}
\hline
  \multicolumn{4}{c}{$\beta$ estimated by Eq.~\eqref{eq::estimate_beta}} \\ \hline
\textbf{}     & it. $1$  & it. $2$   & it. $3$  & it. $4$  \\ \hline
$5 \mu$s & {\color{red} $-0.68$} & $0.72$  & $0.92$ & {\color{red} $-1.01$}  \\ \hline
$20 \mu$s & $0.71$ & $1.10$  & { $4.25$} & { $1.81$}             \\ \hline
$200 \mu$s & {\color{red} $-1.93$} & $1.02$  & { $1.86$} & { $1.99$}               \\ \hline
$2000 \mu$s & {\color{red} $-29.23$} & $0.63$  & $0.69$ & {\color{red} $-2.43$}            \\ \hline
\end{tabular}
\caption{$\beta$ estimation from droplet data, $2048$ physical quantum bits, for various anneling times. Item indices are instance ordinal numbers. Entries in red represent nonphysical (negative) values of the estimated temperature. }\label{tab::beta_droplet}
\end{table}
\begin{table}[ht]
\centering
\begin{tabular}{lcccc}
\hline
  \multicolumn{4}{c}{$\beta$ estimated by Eq.~\eqref{eq::estimate_beta}} \\ \hline
\textbf{}     & it. $1$  & it. $2$   & it. $3$  & it. $4$  \\ \hline
$5 \mu$s & $0.904$ & { $6.64$}  & {\color{red} $-0.49$} &  $1.36$  \\ \hline
$20 \mu$s & $0.76$ & { $2.12$} & {\color{red} $-0.74$} & { $1.47$}             \\ \hline
$200 \mu$s & $0.62$ & { $2.36$} & {\color{red} $-1.78$} & { $1.71$}               \\ \hline
\end{tabular}
\caption{$\beta$ estimation from droplet data, $128$ physical quantum bits, for various annealing times. 
Item indices are instance ordinal numbers. The interpretation is exactly the same as that of Table~\ref{tab::beta_droplet}.}\label{tab::beta_droplet_small}
\end{table}

\section{Conclusions}\label{sec::conclusions}

In this paper, we have introduced an easily implementable method that uses univariate cumulants of order $1$ -- $3$ to assess the quality of the annealer's output. The method results in a parameter, the $p$-value, which under assumptions about the physical model, indicates that a lower part of the energy spectrum has been sampled.

The model depends on the scaling parameter $\alpha$. We use an ad-hoc value which is plausible according to results in the literature for demonstration. Our sensitivity analysis suggests that the model can be moderately sensitive to $\alpha$ in some cases. Yet, its accurate estimation for the given instance (or set of instances) could improve it. We have applied Bootstrap resampling, a heuristic method to compute the significance interval. The method could also be further improved, e.g., by rescaling the significance interval to fit the variance from the error calculus. We have demonstrated the potential of statistical analysis in estimating the quality of Ising annealers' output on particular examples. At least based on the particular examples of annealer outputs we have studied so far, we argue that the introduced analysis can serve as a useful tool in evaluating the solution. To make a stronger statement, the limitations of our model have to be studied further, both analytically (by elaborating, e.g. on the more precise determination of $\alpha$) and empirically by applying the method to many samples. We plan to use it for additional problems in the field of logistics and operations research, similarly to those in ~\cite{grozea2021optimising, salehi2021unconstrained}.

As for the possible further steps of this research, recall that in the case of non-Gaussian distributions (like that of the Ising annealer output), information about a probabilistic model can also appear in cumulants of order higher than $3$. Hence, applying such cumulants may improve our model. Further, as the annealer output is multivariate (containing energy and spin configurations), the multivariate cumulant analysis can give a further clue in analyzing, qualifying, and perhaps correcting the annealer's output. As of the further research, one can also perform quantum annealing with {\tt anneal\_schedule} option of D-Wave, c.f. ~\cite{marshall2019power}. Another interesting question is the systematic study of the effect of the distribution. Systematic testing using artificial data generated assuming non-Boltzmann distribution and assessing the performance of the method would enable the testing of the assumptions in Section~\ref{sec::background} as well as the characterization of the exact limitations. This will be the subject of further research. 

\section*{Acknowledgments} 
The research was supported by the Foundation for Polish Science (FNP) under grant number TEAM NET POIR.04.04.00-00-17C1/18-00 (KD, ZP).
MK acknowledges the support of the National Research,
Development and Innovation Office of Hungary under project numbers K133882 and K124351. 
This research was supported by the Ministry of Innovation and
Technology and the National Research, Development and Innovation
Office within the Quantum Information National Laboratory of Hungary.

We would like to thank Bartłomiej Gardas and Łukasz Pawela for valuable tips on physical models and statistical analysis. We thank Özlem Salehi for suggesting valuable references and technical assistance. We thank Konrad Jałowiecki for technical assistance. We thank the authors of~\cite{rams2018heuristic} for supplying the Droplet instances and their D-Wave solutions. We would like to thank the anonymous referee in pointing out an important error in the previous version of the manuscript.

All data presented on figures are available in the repository \cite{statisticalcode} (sub-directory examples). Other data will be made available by authors under reasonable demand.

\bibliographystyle{ieeetr}
\bibliography{qubo}

\appendix 

\section{Metropolis-Hastings approach for sampling}
\label{sec:MH}

One of the presented samples originates from a simulation of an Ising-based annealer with a  Metropolis-Hastings approach~\cite{chib1995understanding} to sample from the energy spectrum. We call these data artificial as they do not come from a physical annealer. The actual Metropolis-Hastings sampling was performed as follows.

Let $\mathbf{s}_{k}$ be the configuration of spins at step $k$, and consider its $i$-th element.
Following~\cite{chib1995understanding} let $x = \mathbf{s}_{k}$ be a current solution and $y$ be equal to $\mathbf{s}_{k}$, but with a spin flipped at position $i$. Then the probability to move is:
\begin{equation}
    \alpha_{\beta}(x,y) =  \begin{cases}
    \exp(- \beta_{\text{MH}} (H(y) - H(x)) )  \ &\text{if}  \ \ H(y) \geq H(x) \\ 1  \ &\text{if} \ H(y) < H(x),
    \end{cases}
\end{equation}
where the Gibbs distribution with $\beta_{\text{MH}}$ parameter is used -- this parameter models the temperature of the Ising system under simulation. If $\beta_{\text{MH}}$ is low, jumps to ``better'' solutions are favorable, while if the $\beta$ is large, a more extensive search of the solution space is favorable. 
The term $\beta_{\text{MH}} (H(y) - H(x))$ in the first line sets a certain unit to $\beta_{\text{MH}}$. As we have no Boltzmann constant,  the unit of $\beta_{\text{MH}}$ is the inverse of the unit of energy in $H$.

\end{document}